\documentclass[twocolumn,10pt]{article}

\usepackage{amssymb,amsmath,hyperref}

\newcommand\e{{\rm e}}

\newcommand\D{\Delta}

\newcommand\G{\Gamma}
\newcommand\g{\gamma}

\renewcommand\L{\Lambda}
\newcommand\Lt{\tilde \Lambda}
\newcommand\p{\psi}
\renewcommand\P{\Psi}
\newcommand\m{\mu}
\newcommand\n{\nu}
\newcommand\w{\omega}
\newcommand\be{\begin{equation}}
\newcommand\ee{\end{equation}}
\newcommand\bea{\begin{eqnarray}}
\newcommand\eea{\end{eqnarray}}
\newcommand\diag[1]{{\rm diag}(#1)}
\newcommand\tr{{\rm tr}}
\newcommand\x{\ensuremath{\times}}
\newcommand\0{\nonumber} 
\newcommand\C{\ensuremath{\mathbb C}}
\newcommand\q{\ensuremath{{\scriptstyle\,\wedge\,}}}

\newcommand\sect[1]{\medskip\noindent\textsl{\bfseries #1} }

\begin{document}

\newbox\absbox \setbox\absbox\vbox{\noindent\hsize=.95\textwidth
\noindent\textsl{\bfseries Abstract.}  We propose to unify the Gravity
and Standard Model gauge groups by using algebraic spinors of the
standard four-dimensional Clifford algebra, in left-right symmetric
fashion.  This generates exactly a Standard Model family of fermions,
and a Pati-Salam unification group emerges, at the Planck scale, where
(chiral) self-dual gravity decouples. As a remnant of the unification,
isospin-triplets spin-two particles may naturally appear at the weak
scale, providing a striking signal at the LHC.}

\title{Standard Model and Gravity from Spinors} 

\author{Fabrizio Nesti\\
\small\em University of L'Aquila \& INFN -- LNGS,  I-67100,
  L'Aquila\\[2.7em]
\noindent\box\absbox\\[-.5em]
}

\date{}
%

\maketitle



The set of quantum numbers of the fermions in a family of the Standard
Model (SM) is one of the crucial hints that Nature has given us for
understanding the fundamental interactions. While this hint has been
extensively used to guess the fundamental symmetries, it is not
excluded that it could still guide us to surprisingly new structures.
A traditional approach has been to embed a SM family in partial
unification groups like the Left-Right symmetric one
SU(2)$_L\times$SU(2)$_R\times U(1)_{B-L}\times SU(3)$ and the
Pati-Salam one SU(2)$_L\times$SU(2)$_R\times SU(4)$~\cite{LR,PS}, or
into Grand-Unification ones like SU(5), SO(10).
All these approaches consider the gauge groups as internal symmetries,
direct product with the spacetime Lorentz symmetry, and spinors appear
in multiplets of the gauge group.

On the other hand, there is a different way to group fermions in
multiplets, appearing in algebraic spinor theory~\cite{algspin}.  This
is based on the fact that the Clifford algebra is isomorphic to the
algebra of inhomogeneous differential forms, i.e.\ combinations of
zero, one,~\ldots, up to $d$-forms, so that spinors, built out of
these objects, have algebraic as well as geometrical meaning.  While
these spinors still satisfy the Dirac equation, they are not in the
minimal representation of the Lorentz group: they are generic elements
of the Clifford algebra, and thus objects of dimension $2^d$.  As
such, they contain naturally more particles, that appear in multiplets
and can accommodate various sets of quantum numbers, including the SM
ones.

There is a fairly long history of such attempts~\cite{Woit82,
ChisFar87, BayTr92, Bud03, Pav04}, and all provide some evidence of
the emergence of the SM gauge group. In addition they often gauge
together the internal and Lorentz symmetries, and thus represent a
promising setup where also gravity could be reformulated and unified
with the other interactions.

However, most of these approaches end up in postulating extra
dimensions. The reason for this is basically that one SM family
contains 16 Weyl spinors, that is 32 complex components, and thus
$2^4=16$ is not enough to describe even one family. A larger algebra
is necessary, and while it is always possible to enlarge an internal
symmetry, to maintain contact with spacetime geometry this requires
extra dimensions. By doing so, one can easily accommodate a family,
and even motivate the emergence of more families (an even number
usually). This was the strategy pursued in~\cite{ChisFar87}
($Cl_{2,6}$),~\cite{BayTr92} ($Cl_7$),~\cite{Bud03} ($Cl_{9,1}$).

In this letter we suggest that within this framework one can avoid the
use of extra dimensions and accommodate a whole SM family in a
left-right symmetric approach, by using the standard Clifford algebra
of four-dimensional gamma matrices.  Breaking the large underlying
symmetry group leads to a form of Pati-Salam (PS) unification, and the
SM gauge group can then be reached by a standard breaking chain.

The main outcomes of this approach are that: (i)~the existence of the
color SU(4) and SU(2)$_L$\x SU(2)$_R$ groups is connected with the
fact that spacetime is four-dimensional; (ii)~gravity is unified with
the weak isospin groups in chiral way, so that independent selfdual
spin-connections appear; (iii)~the reality of spacetime leads to a
unique Lorentz group while the two weak groups remain split in left
and right copies; (iv)~the metric and its signature emerge from the
symmetry breaking, and (v)~this breaking also predicts a spin-two
isospin-triplet field that may have natural mass at the electroweak
scale;

\sect{Algebraic Spinors.}  We will mainly be concerned with flat space
$R^{1,3}$, and use the standard Clifford algebra $Cl_{1,3}$ given by
Dirac gamma matrices, $\{\g_\m,\g_\n\}=2\eta_{\m\n}$, in Weyl
representation with $\g_5=\diag{1,1,-1,-1}$, $\eta_{\m\n}={\rm
diag}(-1,1,1,1)$. For $Cl_{1,3}$ we use the basis
$\{\g_{A=1\ldots16}\}=\{\mathbf{1}_4,\, \g_0,\, \g_i,\,\g_0\g_i,\,\-
\g_i\g_j,\, \g_5\g_0,\,\g_5\g_i\,,\g_5\}$, with $i,j=1,2,3$.  The real
$Cl_{1,3}$ does not contain the imaginary unit, so it can be taken
over the complex space. Then it is also the algebra $gl(4,\C)$ of all
4$\times4$ complex matrices, and as such it contains different
subalgebras, generically non commuting. For instance one finds
$sl(4,\C)\supset sl(3,\C)\supset sl(2,\C)$ or $sl(4,\C)\supset
sl(2,\C)\times sl(2,\C)$.


While usual spinors are column objects transforming from the left
under Lorentz transformations, algebraic spinors are objects
that themselves belong to the Clifford algebra, $\P=\p^A\,\g_A$, with
$\p^A\in\C$, and transform from the left under algebra-valued
transformations. They can be represented as $4\times4$ complex
matrices,
$$
\P=\left(
\begin{array}{llll}
\p_{11}&\p_{12}&\p_{13}&\p_{14}\\
\p_{21}&\p_{22}&\p_{23}&\p_{24}\\
\p_{31}&\p_{32}&\p_{33}&\p_{34}\\
\p_{41}&\p_{42}&\p_{43}&\p_{44}
\end{array}
\right)\quad \mbox{and}\quad \P\to \e^\L\P\,,
$$
where also $\L=\L^{\!A}\g_A$. Note that $\L$ are generic $gl(4,\C)$
transformations, that is a noncompact algebra.

Evidently such transformations act on each column separately,
therefore the four columns inside the algebraic spinor represent four
invariant subspaces (the four left ideals of the algebra). If we give
to $\m$ the standard meaning of spacetime index, and restrict to a
Lorentz transformation $\L=i\w^{\m\n}\g_\m\g_\n$, we can see that we must
identify each column with a Dirac spinor, and in the Weyl
representation of gamma matrices, the upper and lower halves of each
column are \emph{left} and \emph{right} Weyl spinors.

A further useful property of algebraic spinors is that they naturally
admit transformations also from the right; in general thus:
$$
\P\to \e^\L\P\e^{-\Lt}\,,\qquad \L=\L_A\g_A\,,\qquad \Lt=\Lt_A\g_A\,,
$$
and the $\Lt$ transformations recombine the four columns among them,
commuting with transformations from the left.  They belong to an other
$\widetilde{gl}(4,\C)$, that can accommodate an internal symmetry up to
rank-four, for example~U(4).
Summarizing, an algebraic spinor contains 4 Dirac spinors and these
can be related by some internal symmetry of rank at most four.

Unfortunately, a SM family contains 8 Dirac spinors, therefore this
algebraic spinor is too small. Moreover, also the internal symmetry is
too small to accommodate the SM group in unified way. One easy way out
is to introduce extra dimensions, enlarging the Clifford algebra by
successive factors of two. Naturally this enlarges also both the
transformations from the left and from the right, introducing larger
symmetry groups and extra particles.

Instead of following this approach, we suggest here to give up a
different assumption, namely that the Clifford algebra transformations
$\L$ act on both chiralities, or in other words that the spinor $\P$
contains Dirac spinors in its columns.  We will then introduce two
spinors of opposite chiralities, and each one will be subject to its
own transformations.

\sect{Left and Right Spinors.}  The main
point is that a Clifford transformation contains two commuting sets as
$gl(4,\C)\supset sl(2,\C)\times sl(2,\C)$, that usually are the two
complex-conjugate copies of Lorentz, for the left and right
chiralities.  If we restrict to \emph{left} chirality objects, we can
choose to assign different meanings to them~\cite{graviweak}: one can
be used for one chiral copy of Lorentz, and the other for the weak
isospin, $sl(2,\C)_{lorentz L}\x su(2)_{weak L}$. (One can think
$sl(2,\C)$ and $su(2)$ as having the same generators, $su(2)$
restricted to real parameters.)

We then introduce a \emph{left} algebraic spinor $\P_L$, that is again
algebra valued, $\P_L=\p_L^A\,\g_A$ with complex entries, and that
again transforms from the left under algebra transformations: $\P_L\to
{\rm e}^{\L_L}\P_L$. When $\P_L$ is represented as a $4\times4$
complex matrix, we have again that ${\rm e}^{\L_L}$ acts on each
column separately. However, the $sl(2,\C)_{lorentz L}$ inside ${\rm
e}^{\L_L}$ acts only on half column, so each column contains two Weyl
spinors.  These two spinors are mixed by the commuting $su(2)_{weak}$,
so that inside each column we actually find an isospin doublet of Weyl
fermions.\footnote{Since now Lorentz is only half of $gl(4,\C)$, we
depart from the standard behavior of $\g_\m$ as spacetime vectors, and
we could as well have started with the euclidean $Cl_4$. Indeed the
Lorentz generators (\ref{eq:sl2gen}) are $i\epsilon^{ijk}\g_j\g_k$,
and not $\g_\m\g_\n$.}

As before, $\P_L$ can be transformed also with independent
transformations $\Lt_L=\Lt_L^{A}\g_A$ from the right:
\be
\P_L\to \e^{\L_L}\P_L\e^{-\Lt_L}\,,
\ee
and $\Lt_L\in \widetilde{gl}(4,\C)$ act as internal symmetries, of rank
at most four, e.g.\ U(4)$_L$.

Summarizing, $\P_L$ contains 4 isospin doublets of Weyl spinors, that
we can now identify with the \emph{left} half of a standard model
family. Moreover, since they are related by an internal symmetry of
rank 4, it is very suggestive to represent $\P_L$ also with lepton and
colored quarks indices:%
\vspace*{-.5ex}
\be
\label{eq:psidef}
\P_L=\left(
\begin{array}{llll}
 \nu _{\text{L1}} & u_{\text{L1},r} &
   u_{\text{L1},g} & u_{\text{L1},b}
   \\
 \nu _{\text{L2}} & u_{\text{L2},r} &
   u_{\text{L2},g} & u_{\text{L2},b}
   \\
 e_{\text{L1}} & d_{\text{L1},r} &
   d_{\text{L1},g} & d_{\text{L1},b}
   \\
 e_{\text{L2}} & d_{\text{L2},r} &
   d_{\text{L2},g} & d_{\text{L2},b}
\end{array}
\right).
\vspace*{-.5ex}
\ee
A SM family is described just by a left-right symmetric couple of such
spinors: $\P_L$, $\P_R$.

\medskip

Let us now discuss the (global) internal symmetries. First we observe
that since we speak now of objects of separate chirality, at this
stage everything is duplicated.  From the independent $\L_L$, $\Lt_L$,
$\L_R$, $\Lt_R$, we have
\be
\label{eq:gengroup}
gl(4,\C)_{L}\times \widetilde{gl}(4,\C)_{L}\ \times\ 
gl(4,\C)_{R}\times \widetilde{gl}(4,\C)_{R}\,.
\ee
One should think these as geometric symmetries, dictating the field
representations, only partially realized in our low energy world.
This is exactly what we will find when trying to write a theory in
spacetime.

Let's restrict first to one chirality. The $\L_{L}$ transformations,
belonging to $gl(4,\C)$, contain generators that mix Lorentz and weak
indices as well as noncompact ones, that are not observed in our
world.  We will thus need a breaking of $gl(4,\C)_L$ to
$sl(2,\C)_{lorentz L}\x su(2)_{weak L}$. 

The $\Lt_L$ transformations belong to an other $\widetilde{gl}(4,\C)$
and should be restricted to be compact, because noncompact internal
symmetries are always plagued by ghosts. This minimal requirement
leads from $\widetilde{gl}(4,\C)$ to its maximal compact group U(4),
i.e.\ a group that unifies color and $B-L$ by treating lepton as the
fourth color, as in the celebrated Pati-Salam group.  The
representation (\ref{eq:psidef}) of $\P_{L}$ explicitly showed
this.

 In this ``broken'' phase the symmetries would thus be
$sl(2,\C)_{lorentzL}\times su(2)_{weakL}\times u(4)_{L}$.  Then, the
\emph{right} chirality would in principle give rise to a second copy
of these, i.e.\ $sl(2,\C)_{lorentzR}\times su(2)_{weakR}\times
u(4)_{R}$.  If on one hand it is nice to have a duplication of the
weak isospin group because the SM quantum numbers suggest it, on the
other hand especially the doubling of Lorentz should be removed in the
real world.

Before describing the broken phase where this happens, it is useful to
lay down explicitly the generators of $sl(2,\C)_{lorentz}\times
su(2)_{weak}$ inside $\L_{L,R}$:
\bea
\nonumber
\!\!su(2)_{weak}~~\!:&&\!\!\!\!\!\{\tau_{i}\}
=\{i\g_0,-\g_0\g_5,\g_5\}\!\equiv\!\{\mathbf1_2\otimes\sigma_i\}\\[.4ex]
\label{eq:su2gen}
\label{eq:sl2gen}
\!\!\!\!\!\!\!\!\!\!\!sl(2,\C)_{lorentz}\!:&&\!\!\!\!\!\{\rho_{i}\}
=\{-i\epsilon_{ijk}\g_j\g_k\}\!\equiv\!\{\sigma_i\otimes\mathbf1_2\}\,.
\eea

\sect{Broken phase from fermion kinetic terms.} 
To describe propagating fermions, one needs to introduce the
equivalent of spacetime gamma matrices, i.e.\ a vierbein or soldering
form $\G_\m$ connecting fermion bilinears with the spacetime
derivative $\partial_\m$. $\G_\m$ is an algebra-valued field that
transforms as $\G_\m\to \e^{-\L^\dagger}\G_\m\e^{-\L}$ under
$gl(4,\C)$, and can be taken of 16 real components (for each index
$\m$).

While a generic $\G_\m$ would be needed in a symmetric phase, we
assume that it develops a VEV $\bar \G_\m$, defining the broken phase
where we live. To have Minkowski spacetime, $\bar \G_\m$ should leave
unbroken a global $so(1,3)_{lorentz}$ symmetry defined as simultaneous
transformations $\L_\m^\n$ of 1)~a Lorentz subgroup of spacetime
diffeomorphisms and 2)~internal $\L$ transformations restricted to
$sl(2,\C)_{lorentz}$: $\bar\G_\m\equiv
\e^{-\L^\dagger}\bar\G_\n\e^{-\L} \L_\m^\n $, as it happens for
ordinary gamma matrices.  This defines a soldering of the spacetime
and internal Lorentz symmetry groups, and thus the signature of
spacetime emerges from the VEV $\bar\G_\m$.

Two different $\G_\m^{L,R}$ should actually be defined, that do this
job for the two $sl(2,\C)_{lorentz L,R}$, and their VEVs should also
be aligned as $\bar \G_\m^L=\eta_{\m\m}\bar \G^R_\m$ so that they
define the same Minkowski metric $\eta_{\m\n}$ for L and R spinors.
Explicitly:
\be
\label{eq:VEV}
\bar \G^{L,R}_\m = \{\pm \mathbf{1}_4,-i \epsilon_{ijk}\g_j\g_k\}
                 = \{\pm\mathbf{1}_2,\sigma_i\}\otimes\mathbf{1}_2\,.
\ee
$\bar\G_\m^{L,R}$, $su(2)_{weak L,R}$ commute as one checks with
(\ref{eq:su2gen}).

We can now build kinetic terms for \emph{left} and \emph{right}
fermions ($\tr$ is the trace in 4$\times$4
representation):\footnote{For the curved and first-order action in the
symmetric phase, an inverse soldering should be defined,
see~\cite{graviweak,future}.}
\vspace*{-.5ex}
\be
\label{eq:lkin}
{\cal L}=\tr[\P_L^\dagger\partial^\m \bar\G^L_\m\P_L]+\tr[\P_R^\dagger\partial^\m \bar\G^R_\m\P_R]\,,
\ee
and look for the remaining global invariances.

These kinetic terms restrict some subgroups to have unitary elements
${\rm e}^\L{\rm e}^{\L^\dagger}=\mathbf 1$, i.e.\ to be compact.  This
happens to the $sl(2,\C)_{weak}$ groups, whose generators commute with
$\bar\G_\m$, and also to the $\widetilde{gl}(4,\C)$.  Therefore these
are reduced respectively to the compact groups $su(2)_{weak}$ and
$u(4)=su(4)\times u(1)$.\footnote{The breaking $gl(4,\C)\to u(4)$ may
be realized e.g.\ with a field transforming as
$\Phi\to\e^{\Lt}\Phi\e^{\Lt^\dagger}$, with VEV
$\bar{\Phi}=\mathbf1_4$.}
The $sl(2,\C)_{lorentz}$ groups on the other hand follow a different
fate: they remain non compact because a generic transformation, not
necessarily unitary, is compensated by a spacetime Lorentz
transformation of $\partial_\m$; indeed the internal Lorentz is
soldered with the spacetime one.  In addition, since the derivative
$\partial_\mu$ is the same in both the L and R sectors, then
$sl(2,\C)_{lorentz L,R}$ are actually forced to transform together
\be
\label{eq:reality}
\L_{lorentz L}^\dagger=-\L_{lorentz R}
\ee
and we remain with just one diagonal $sl(2,\C)_{lorentz}$, identified
with the spacetime Lorentz group $so(1,3)$.

The mixed Lorentz-weak transformations do not preserve $\bar\G_\m$ and
are thus broken.


Summarizing, in the broken phase, in place of the large non compact
group (\ref{eq:gengroup}) we have (at most) the symmetry:
$$
so(1,3)_{lorentz}\times su(2)_{L}\times su(2)_{R}\times u(4)_L\times
u(4)_R\,.
$$ 
This group can be linked to the Pati-Salam and SM groups by standard
breaking chains, after introducing appropriate Higgs fields needed for
the symmetry breakings.  In particular, $su(4)_L\times
su(4)_R\to\,su(4)\,\to\,su(3)_{color} \times u(1)_{B-L}$, can be
realized by a field $\Phi_{\tilde L\tilde R}\in(\mathbf{4}_{\tilde
L},\mathbf{4}_{\tilde R})$ of
$\widetilde{gl}(4,\C)_L\times\widetilde{gl}(4,\C)_R$, i.e.\
transforming as $\Phi_{\tilde L\tilde R}\to\e^{\Lt_L}\Phi_{\tilde
L\tilde R}\e^{-\Lt_R}$, with VEV $\bar \Phi_{\tilde L\tilde
R}=\diag{-3,1,1,1}$.  The electroweak breaking leads to a similar
field: since here the Lorentz and weak groups are unified, it is not
possible to find directly a multiplet containing the scalar Higgs
doublet. In fact isospin doublets are here also lorentz doublets,
i.e. spinors, and are contained in the vector representation
$\mathbf{4}$ of $gl(4,\C)$.  The solution is found by coupling the L
and R sectors (and this is in a sense natural) with a field
$\Phi_{LR}\in(\mathbf{4}_L,\mathbf{4}_R)$ that transforms as a vector
under both graviweak groups $gl(4,\C)_L\times gl(4,\C)_R$, i.e.\
$\Phi_{LR}\to\e^{\L_L}\Phi_{LR}\e^{-\L_R}$. Since in the broken phase
the two Lorentz subgroups are soldered, this decomposes as
$(\mathbf1+\mathbf3,\mathbf2_L,\mathbf2_R)$ of $so(1,3) \times
su(2)_L\times su(2)_R$, showing nicely the emergence of the scalar
bidoublet $(\mathbf1,\mathbf2_L,\mathbf2_R)$.

As in left-right symmetric theories, the L and R weak groups may be
broken at different scales giving rise to the observed parity-breaking
phenomenology~\cite{LRbreaking}. Of course, a complete model should
also include the mixing matrices, as well as the mechanism for the
quark/lepton, U/D and horizontal hierarchies of masses.  All these
aspects may require additional fields, and will then be constrained by
the unified dynamics beyond the planck scale.  We leave the full
analysis for future model building~\cite{future}, and proceed to
discuss the gauge fields.

\sect{Gauging.}  The gauging of symmetries inside $\L$ and $\Lt$ is
realized by introducing a covariant derivative with Clifford-algebra
valued vector fields, each having 16 complex components:
\be
\label{eq:covariantD}
\partial_\m  \rightarrow  
D_\m^{L,R}=\partial_\m +V_\m^{L,R}+\tilde V_\m^{L,R}\,,
\ee
with $V_\m$, $\tilde V_\m$ acting from the correct sides of the
fields.

While the tilded fields $\tilde V_\m^{L,R}$ are just the gauge fields
of (complexified) $u(4)_{L,R}$ in Clifford algebra notation, the
fields $V_\m^{L,R}$ are more interesting: they unify gravity and
weak-isospin, that are contained in the $gl(4,\C)$ algebras.

It must be noted that in $V$ and $\tilde V$ one may prefer avoid
gauging the $u(1)$ factors (the identity $\mathbf 1_4$): in fact the
first is a gauging of dilatations that usually leads to unimodular
gravity (eating the trace of the graviton, see below) and the second
is anomalous.  Therefore we will start from a gauging of four copies
of $sl(4,\C)$.  With this choice, in the broken phase the only
non-decoupled gauge fields are the $su(4)$ and the $su(2)$ ones.  To
be explicit the $V^{L}_\m$ can be parametrized in terms of the complex
gauge fields $\w^L$, $\hat W^L$, $Z^L$ (and similarly for $V^R_\m$):
\be
V^L_\m=i\w_\m^{L\,i}(\sigma_i\otimes\mathbf1_2) + i \hat W_\m^{L\,i}(\mathbf1_2\otimes\sigma_i)+Z_\m^{L\,ij}(\sigma_i\otimes\sigma_j).
\ee
In a flat broken phase, $V_{L,R}$ will then contain only the
$su(2)_{L,R}$ gauge fields, i.e.\ the (real) $W^{L,R}_\m$ gauge
bosons, while in the curved case also a component along the
$sl(2,\C)_{lorentz}$ generators $\rho_i$ is present:
\bea 
V_\mu^{L}\!& =&\! i \w_\m^i \,\rho_i +iW_{\m}^{R\,i}\, \tau_i\0\\ 
V_\mu^{R}\!& =&\! i \bar \w_\m^i \, \rho_i +iW_{\m}^{L\,i}\, \tau_i\,,\\[-1.5em]\0
\eea 
%
The $\w_\m^i$, $\bar \w_\m^i$ reproduce the selfdual and antiselfdual
spin connections of gravity, conjugate acting on the left or right
fields, as in the reality condition (\ref{eq:reality}). In fact one
can make contact with Einstein gravity by recalling that $\w_\m^i$ is
expressed in terms of the Christoffel symbols as $\w_\m^i =
\epsilon^i_{jk}\G_\m^{jk}+i\G_\m^{0i}$, that is also the selfdual part
of a \emph{complex} Christoffel connection (see
e.g. \cite{chams,peldan,graviweak}). We expect this to happen based
simply on symmetry arguments, since as discussed above, the two
internal Lorentz groups are broken (to the global $so(1,3)_{lorentz}$)
therefore also the fluctuations around $\w_\m^i$, $\bar\w_\m^i$ are
massive and decoupled. 

This also tells us that the VEVs $\bar\G_\m$ are fixed to be at
$M_{pl}$, in the Palatini spirit.  Indeed, in the broken phase, the
existence of $\bar\G_\m$ allows to write a linear Einstein-Hilbert
action ${\cal L}_{EH}=\tr(R\/\/\bar\G\bar\G)$ only for the
gravitational curvature $R_{\m\n}$ (and not for the weak gauge
field-strength) defining the dimensionful gravitational coupling
$M_{pl}^2\sim\bar \G^2$.  

On the other hand a quadratic action can be written for both
curvatures, $R_{\m\n}^2$, $W_{\m\n}^2$, whose couplings will be
unified in the symmetric phase~\cite{graviweak}.

\sect{Fluctuations.}  
Since $\G_\m^{L,R}$ act as 'Higgs' fields, it is interesting to
analyze their fluctuations.  Each $\G_\m^{L,R}$ has 64 components,
that can be decomposed as:
$$ 
\G_\m= M_{pl}(\eta_{\m\n}+h_{\m\n})(\hat\sigma^\m\otimes\mathbf{1}_2) +
\D_{\m\n}^i(\hat\sigma^\m\otimes\sigma_i)
$$
where $\hat\sigma^\m=\{\mp\mathbf{1}_2,\sigma_i\}$ for $L$, $R$. The
first term is the backgound value defining here minkowski space, and
$h$ and $\D$ are the (real) fluctuations.  It is straightforward to
identify among them the goldstone fields of the $gl(4,\C)$ symmetry
breaking, that in the unitary gauge are 'eaten' by the corresponding
gauge fields. These are: the antisymmetric part of $h_{\m\n}$, eaten
by the fluctuations of the spin connection $\w$; the three
antysimmetric parts of $\D_{\m\n}^i$, eaten by the fields mixing
lorentz and weak symmetry $Z^{ij}_\m$; and finally the three traces
$\D_{\m}^{\m\,i}$ that give mass to the noncompact (imaginary) isospin
gauge fields, ${\rm Im}\, \hat W^i_\m$.

Summarizing, after symmetry breaking we find the following low energy
field content: two standard gravitons, $h_{\m\n}^L$, $h_{\m\n}^R$ (10
components each) and two new traceless gravitons $\D_{\m\n}^{L\,i}$,
$\D_{\m\n}^{R\,i}$ that are isospin-triplets for the respective
$su(2)_{L,R}$ groups (27 components each).\footnote{The appearance of
'colored' gravitons was described in~\cite{isham,chams2}.  Here however
breaking $gl(4,\C)$ implies that also the antisymmetric part of $\D$
is eaten and is thus completely decoupled.}

\sect{Phenomenology.}
Let us discuss the phenomenology of these fields.  Of the two
singlet-gravitons, a parity-even combination $h^+=h^L+h^R$ will be
massless due to diffeomorphism invariance, and will correspond to the
standard graviton.  It will also couple universally to L and R matter.
The parity odd combination $h^-=h^L-h^R$ on the other hand is not
protected by diffeomorphisms and may be massive
(see~\cite{isham,bigr}). Its mass can not be predicted at this stage,
because it depends on the details of the full theory at energy beyond
the planck scale. It is clear that if $h^-$ had planck-scale mass it
would be unobservable at low energy, while if it were sufficiently
light it would give rise to polarization-dependent gravitational
effects, among these gravitational waves.

A different situation arises for the $L$, $R$ triplet-gravitons: they
are not protected by diffeomorphisms therefore they both can be
massive.  It is then interesting to observe that each $\D$ can have
two kinds of mass terms: one is 'gauge invariant', and the other comes
from a coupling with e.g.\ one doublet higgs $\phi$ responsible for the
SU(2) symmetry breaking. Setting $\D_{\m\n}=\D_{\m\n}^i\sigma_i$:
\vspace*{-.5ex}
\be
\tr(\D_{\m\n})^2\,,\qquad |\D_{\m\n}\phi|^2\,.
\vspace*{-.5ex}
\ee
Now, the first one is originated from terms of the form $\G^4$, that
give rise also to the cosmological constant (see
e.g.~\cite{graviweak,chams2} for explicit constructions) therefore one
may expect it to be small, with the result that the mass of the
$\D_{\m\n}^{i\,L,R}$ is linked to the scale of breaking of the
relative SU(2)$_{L,R}$ group.

The $L$ triplet is then particularly interesting, because it may have
natural mass in the weak range, and being charged under isospin, would
be easily produced and observed at the LHC (while its other couplings
to matter are gravitational, i.e.\ planck suppressed). The present
experimental lower bound on its mass can be estimated to be as low as
$\sim$300~GeV, from Drell-Yan pairwise production at Tevatron. It
would be interesting to estimate the background and thus the
corresponding discovery reach of LHC.

As already remarked, the precise mass predictions depend on a complete
formulation of the symmetric phase, that we can address only partially
here.

\sect{Symmetric phase.}
The discussion relied up to now on the use of the extended vierbeins
$\G_\m^{L,R}$, that were assumed to develop a VEV, spontaneously
breaking the large symmetry~(\ref{eq:gengroup}).  It is natural to ask
whether a theory in the fully symmetric phase could be given to
explain such a breaking.  The question was answered affirmatively in
models of 'graviweak' unification~\cite{graviweak} with smaller,
orthogonal gauge groups such as $SO(4,\C)$ or $SO(7,\C)$, also using
an extended vierbein.  The action in the symmetric phase is simply
${\cal R}\q\G\q\G\epsilon$, where the $\epsilon$ invariant tensor was
derived from the duality operator in the corresponding algebra (acting
in the representation of the $\G$'s), and ${\cal R}$ is the full
curvature of the connection~$V$.

In the present scenario, since $gl(4,\C)$ has no duality, one can not
write an invariant theory using only $\G$ and ${\cal R}$, and at least
one new field has to be introduced.  Indeed one possibility would be
to proceed as in Plebanski-inspired
constructions~\cite{plebanski,smolin}, where exactly the missing
antisymmetric field is introduced.

Also, to understand the predictions in terms of masses of $\D^{L,R}$,
one should introduce the full set of Higgs fields for the electroweak
breaking, or in other words, to complete the full theory.  We expect
that, as is common in grand-unified theories, and unlike the bottom-up
approach described here, such completions will not be unique, and plan
to investigate them in a separate work~\cite{future}.

Some comments can nevertheless be made on the gravitational sector: in
the symmetric phase, prior to symmetry breaking, there is no metric
and necessarily propagation is not standard: for example, if $\G_\m$
has no VEV fermion kinetic terms are not gaussian, and if the theory
is formulated in first order formalism, the gauge action has a single
derivative~\cite{graviweak, future}.  This is a well known fact in the
Weyl-Cartan-Palatini approach to gravity, and it can be related to the
problem of its quantization.  Here one can speculate that some help in
understanding this phase will come from the underlying geometrical
structure.  A glimpse of this possibility can already be seen in the
unusual emergence of the Lorentz group: to use chiral algebraic
spinors, we needed in the symmetric phase two copies of the internal
Lorentz group; then the fact that there is only one \emph{real}
spacetime, forces them to be glued in the broken phase. It is thus the
soldering of $\bar\G_{L,R}$ with a real spacetime that leads to a
unique Lorentz group for spinors.\footnote{ This fact can be
understood in the known theory of SL(2,\C) spinors: SO(1,3) is
(locally) isomorphic to SL(2,\C), and does not factorize as
SL(2,\C)$\times$SL(2,\C); this instead is true for its
complexification, that may act only on a complex spacetime. Indeed,
from a mathematical point of view, doubling the algebraic spinors is
like doubling the (co)tangent space, that may be interpreted as a
complexification of spacetime, along the lines
of~\cite{Woit82,chams,peldan}. It is not clear how to interpret
geometrically and dynamically the section to a real spacetime.}  At
the same time an analogous reality condition does not apply to the
weak groups, that thus remain split in L and R parts, nicely
predicting the left and right isospin fermion doublets.  However, to
achieve these results the theory demanded us not only two different
extended spin connections, but also two extended vierbeins, or in
other words two full (self-dual) gravities living in the L and R
sectors, above the Planck scale.

The appearance of selfdual spin-connections in $V_{L,R}$ is a direct
consequence of the choice of working with chiral spinors; nevertheless
it is also most welcome, since there has been considerable progress in
the formulation of gravity at both the classical and quantum level by
using these variables~\cite{ash}.  They allow a much easier
Hamiltonian formulation, even if there are difficulties related to the
reality conditions.
In the setup described here the L and R sectors are independent at the
fundamental level, and the reality conditions are expected to arise
dynamically (like~(\ref{eq:reality})). We stress indeed that
differently from an enlargement of the gauge group of gravity, as
in~\cite{chakpeld}, here we were led to its duplication in L and R
sectors, and both gauge connections decouple because they `eat' two
independent sets of goldstone bosons in the vierbeins $\G^{L,R}$.

We observe incidentally that $\bar \G_\m^{L,R}$ could in general lead
to a phase with two \emph{different} coexisting background metrics.
In such a phase one could have observable violations of Lorentz
invariance, but also healthy massive gravitons and phenomenologically
interesting consequences~\cite{bigr}.

We finally want to observe that due to the intrinsic geometric nature
of algebraic spinors, it may also be that the symmetric phase
described here is the effective geometric description of a completely
different microscopic theory, where also fermions may be
emergent. Therefore the present work may also be a step forward for a
realistic effective description of matter, where the SM group can
emerge naturally in Left-Right symmetric setup.

\sect{Conclusions.} 
 What we found is that adopting algebraic spinors, and using them in
left-right symmetric fashion, we unified the fermions of a SM family.
The framework is by construction endowed with a large non compact
group, unifying the gravitational and gauge interactions.  Then, the
requirement to have a broken phase with a field theory in spacetime
reduces the internal symmetries to be compact and also
phenomenologically close to the SM, pointing to a Pati-Salam
unification.

We identified the fields responsible for this breaking, as extended
vierbeins whose VEV gives rise both to the metric and to the chiral
weak groups, pointing toward a duplication of the connections, as two
copies of self-dual extended gravities.

We discussed only partially the most fundamental level of the
symmetric $gl(4,\C)$ phase extending beyond the Planck scale, where
the symmetries strongly restrict the representations and actions, and
thus the dynamics leading to the scales of the breakings.  This part
of the analysis implies formulating the theory in first order, curved
space and $gl(4,\C)$-invarient way and will be addressed in a separate
work.

Finally, a possible experimental signature of this unification has
emerged, because the extended vierbeins also contain isospin-triplet
tensor fields, and the left one may naturally have mass at the weak
scale. If this is the case, it would strikingly show up at the LHC as
weakly interacting spin-two particles.

\sect{Acknowledgments.}  I would like to thank M. Pavsic for
discussing algebraic spinors for the SM, and R.~Percacci for useful
discussions on the broken phase.  Also, interesting exchanges with
W.E.~Baylis, G. Trayling and discussions with Z.~Berezhiani,
D. Comelli, L.~Pilo, G.~Senjanovi\'c, are acknowledged.  Work
partially supported by the MIUR grant PRIN-2006 ``Astroparticle
Physics''.
Thanks also to INFN and ICTP for support during the completion of this
work.

\end{document}